\documentclass[12pt]{iopart}
\usepackage{graphicx}
\usepackage{setspace}
\usepackage{texdraw}

\begin{document}
\title{Analysis of  entropy  of $XY$ Spin Chain }

\author{F. Franchini $\diamond$,  \ A. R. Its\dag,  \ B.-Q. Jin\S   \ and V. E. Korepin $\star $ }

\address{ $\diamond$ Department of Physics and Astronomy, State University of New York at \\
Stony Brook, Stony Brook, NY 11794, USA}
\address{ \dag\ Department of Mathematical Sciences, Indiana
University-Purdue University Indianapolis, Indianapolis, IN
46202-3216, USA}
\address{\S\ College of Physics and Electronic Information, Wenzhou
University, Wenzhou, Zhejiang, P.R. China }
\address{$\star $\ C.N.\ Yang Institute for Theoretical Physics, State
University of New York at Stony Brook, Stony Brook, NY 11794-3840,
USA}

\ead{Fabio.Franchini@stonybrook.edu,itsa@math.iupui.edu,jinbq@wzu.edu.cn,
korepin@insti.physics.sunysb.edu}


\begin{abstract}
Entanglement in the ground state of the  $XY$ model on the
infinite chain can be measured by the von Neumann entropy of a
block of neighboring spins. We study a double scaling limit: the
size of the block is much larger then $1$ but much smaller then
the length of the whole chain. In this limit, the entropy of the
block approaches a constant. The limiting  entropy is a  function of the anisotropy and of the
 magnetic field. The entropy reaches  minima at
 product states and  increases boundlessly at
phase transitions.
\end{abstract}

\maketitle

\section{Introduction}

Entanglement is a primary resource for quantum computation and
information processing \cite{BD,L, ben,pop}. It  shows how much quantum effects we can use to
control one system by another. Stable and large scale entanglement
is necessary for scalability of quantum computation
\cite{rasetti,  GRAC}. The entropy of a subsystem as a measure of
entanglement was discovered in \cite{ben}. Essential progress has
been achieved in the understanding of entanglement in various
quantum systems
\cite{rasetti,fazio,jin,LRV,K,julien, cardy,ABV,VMC,LO,PP, eisert2,
briegel, bruno}.

The importance of the $XY$ model for quantum information  was
emphasized in \cite{fazio,vidal,keat,bose}. In this paper we
consider the entropy of a block of $L$ neighboring spins in the
ground state of the $XY$ model [on the infinite chain] in the
limit $L\rightarrow \infty $. We use the results of \cite{bik,
pesh, big}. The Hamiltonian of the $XY$  model is
\begin{equation}
{\cal H}=-\sum_{n=-\infty}^{\infty}
(1+\gamma)\sigma^x_{n}\sigma^x_{n+1}+(1-\gamma)\sigma^y_{n}\sigma^y_{n+1}
+ h\sigma^z_{n} \label{xxh}
\end{equation}
Here $0<\gamma<1$ is the anisotropy parameter; $\sigma^x_n$,
$\sigma^y_n$ and  $\sigma^z_n$ are the Pauli matrices and $0<h$ is
the magnetic field. The model was solved in
\cite{Lieb,mccoy,mccoy2,gallavotti}. The methods of Toeplitz
determinants and integrable Fredholm operators were used for the
evaluation of correlation functions, see  \cite{mccoy2,aban,sla,
dz, izer, pron}. The idea to use the determinants for calculation of entropy was put forward in \cite{jin}.

 Solution of $XY$ looks  differently in {\Large three  cases:} $\Downarrow$

 {\bf Case $1$a} is defined by the inequality $\quad 2\sqrt{1-\gamma^2}<h< 2$ .

It describes moderate magnetic field.

 {\bf Case $2$} is defined by $h> 2$. This is strong magnetic field.

  {\bf Case $1$b} is defined by $0<h<2\sqrt{1-\gamma^2}$.

  It is weak magnetic field, including zero magnetic field.
 \vfill\eject

At the boundary between cases $1$a and $1$b ($  h=2{\sqrt{1-\gamma^2}} $)
 the ground state is doubly degenerated:
\begin{eqnarray}
		  |G_1\rangle & = & \prod_{n\in \mbox{\rm lattice}}\left[ \cos(\theta)| \uparrow _n \rangle+ (-1)^{n}\sin(\theta)| \downarrow _n \rangle \right] , \qquad  \qquad \qquad \bowtie  
  \nonumber   \\
  |G_2\rangle & = &  \prod_{n\in \mbox{\rm lattice}}\left[ \cos(\theta)| \uparrow _n\rangle - (-1)^{n}\sin(\theta)| \downarrow _n \rangle \right] 
  \label{deg}
\end{eqnarray}
Here  $\cos^2 (2\theta) =(1-\gamma)/(1+\gamma)$. The role of
factorized states was emphasized in \cite{aban, ver,tog}.
Let us mention that the rest of energy levels are separated by a gap and 
correlations decay exponentially. The boundary  boundary between cases $1$a and $1$b is not a phase transition.

In general, we denote  the ground state of the model by
$|GS\rangle$. We consider the entropy of  a block  of $L$
neighboring spins: it measures the entanglement between the block
and the rest of the chain \cite{ben,vidal}. We treat the whole
ground state as a binary system $|GS\rangle = |A \& B\rangle $.
The block of $L$ neighboring spins is subsystem A and the rest of
the ground state is subsystem B. The density matrix of the ground
state is \mbox{$\rho_{AB}=|GS\rangle \langle GS|$}. The density
matrix of the block is \mbox{$\rho_A= Tr_B(\rho_{AB})$}. The
entropy $S(\rho_A)$ of the block is:
\begin{equation}
S(\rho_A)=-Tr_A(\rho_A \ln \rho_A)   \qquad \qquad \qquad  \sharp  \label{edif} 
\end{equation}
{\bf Note that each of the ground states (\ref{deg}) is factorized and
has no entropy.}

To express the entropy we need the complete elliptic integral of
the first kind,
\begin{equation}
I(k) = \int_{0}^{1}\frac{dx}{\sqrt{(1-x^2)(1 - k^{2}x^{2})}}  \qquad \qquad \star
\nonumber
\end{equation}
and the modulus
\begin{equation}
   \tau_0= I(k')/I(k), \qquad \qquad k'=\sqrt{1-k^2} 
\end{equation}

In the paper \cite{bik} we used determinant representation for evaluation of entropy. Zeros of the determinant  form an infinite sequence of  numbers:
\begin{equation}\label{zerosMay}
\lambda_{m} =
  \tanh \left(m + \frac{1-\sigma}{2}\right)\pi \tau_{0}, \quad m \geq 0,\qquad 
\lambda_{-m} =-\lambda_{m} 
\end{equation}
here $\sigma = 1$ in Case 1 and $\sigma = 0$ in Case 2.
Note  
 $0 < \lambda_{m} < 1$ and  $ \lambda_{m} \to 1$ as  $ m \to \infty$

The magnetic field and anisotropy define $k$:
\begin{eqnarray}
   k= \left \{ \begin {array} {c} \sqrt{(h/2)^2+\gamma^2-1}\; /\; \gamma ,
  \;\;\; \qquad \qquad  \mbox{Case 1a} \\ [0.3cm]
  \sqrt{1 -\gamma^2 - (h/2)^2}\; / \;\sqrt{1-(h/2)^2},
\;\;\; \mbox{Case 1b}\\ [0.3cm]
         \gamma\; / \;\sqrt{(h/2)^2+\gamma^2-1} ,\;\;\;\qquad \qquad  \mbox{Case 2}
 \end{array}
  \right.
    \label{modMay}
  \end{eqnarray}

We represented the entropy as a convergent series in \cite{bik}:
\begin{equation}\label{3333May}
 \heartsuit  \quad \qquad  S(\rho_A) = 
\sum_{m=-\infty}^{\infty}
  (1+\lambda_{m})\ln \frac{2}{1+\lambda_{m}}  \qquad  \spadesuit
  \end{equation}
I. Peshel using the approach of \cite{cardy} also obtained the series (\ref{3333May})
in cases of  non-zero magnetic field, see  \cite{pesh}.
He summed it up into:
$$
 S(\rho_A)=   \frac {1} {6} \left [\;\ln{ \left (\frac {k^2} {16 k'}\right )} +
\left (1-\frac {k^2} {2}\right )
         \frac {4 I(k) I(k')} {\pi} \right ] + \ln\;2,  \qquad {\mbox{ Case 1a}} $$
$$
   S(\rho_A)=  \frac {1} {12} \left [\;\ln{ \frac {16} {(k^2 k'^2)}} +
(k^2-k'^2)
         \frac {4 I(k) I(k')} {\pi} \right ],   \qquad  \qquad \quad 
{\mbox{ Case 2}} $$
   We summed up the series    (\ref{3333May})  in case of  weak magnetic field (including zero magnetic field) in the paper \cite{bik}:
$$
 \clubsuit \qquad   S(\rho_A) =   \frac {1} {6} \left [\;\ln{ \left (\frac {k^2} {16 k'}\right )} + \left (1-\frac {k^2} {2}\right )
         \frac {4 I(k) I(k')} {\pi} \right ] + \ln\;2,   \qquad \mbox{ Case 1b},  $$
The  rigorous proof  and the precise history is given in the
paper \cite{big}.

\begin{figure}
 \includegraphics[width=\columnwidth]{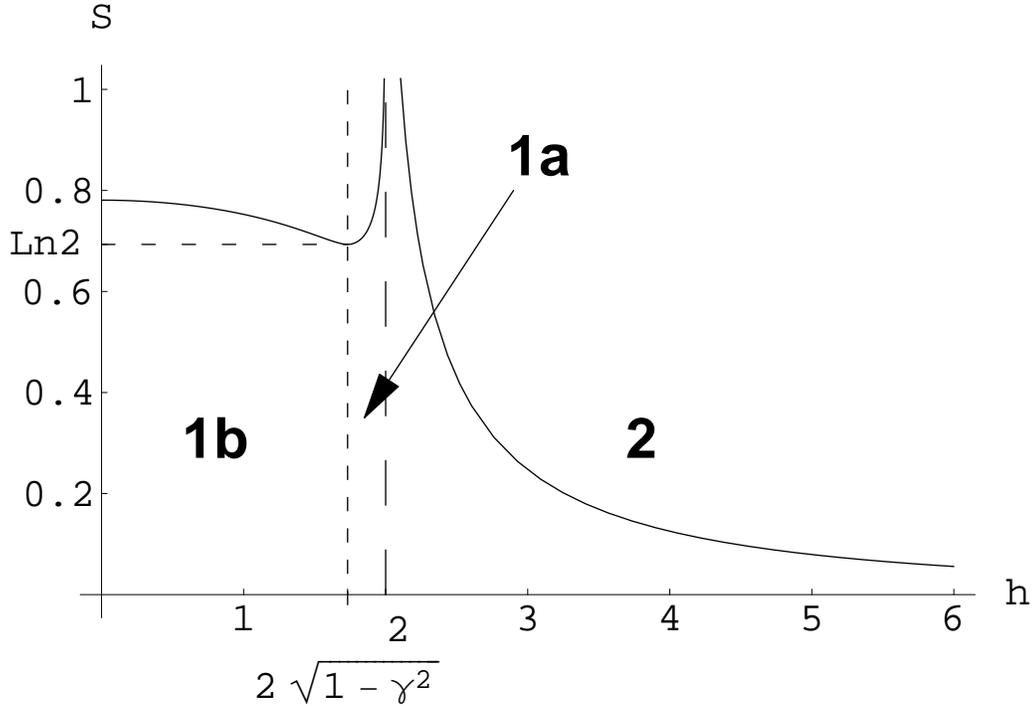}
\caption {The limiting entropy as a function of the magnetic field
at constant anisotropy $\gamma = 1/2$. The entropy has a local minimum $S=\ln2$ at
$h=2\sqrt{1-\gamma^2} $ and the absolute minimum for $h \to \infty$ where it vanishes.
$S$ is singular at the phase transition $h=2$ where it diverges to $+ \infty$.
The three cases are marked.}
   \label{entropyplot}
\end{figure}

Now we  can study the range of variation of the limiting entropy.
We find a {\bf  local minimum} $ S(\rho_A)=\ln2$ at the boundary between
cases $1$a and $1$b ($h=2\sqrt{1-\gamma^2}$). This is the case of
double degenerated ground state (\ref{deg}). The absolute minimum
is achieved at infinite magnetic field, where the ground state
becomes ferromagnetic (i.e. all spins are parallel). The entropy
diverges to $+ \infty$, i.e has singularities, at the {\it phase
transitions}: $h=2$  \cite{cardy} or $\gamma=0$, see \cite{jin}.
To show this behavior of the limiting entropy, we plot it as a
function of the magnetic field $h$ at constant anisotropy $\gamma
= 1/2$ in Fig.~\ref{entropyplot}.
It is interesting to note that the critical behavior of the $XY$
model is similar to the Lipkin-Meshkov-Glick model \cite{julien}.

{\it Acknowledgments.} We would like to thank P.Deift, B.McCoy,
I.Peschel and H.Widom for useful discussions. This work was
supported by NSF Grants DMR-0302758, DMS-0099812 and DMS-0503712.

\end{document}